\documentclass[11pt,russian,a4paper,twoside]{article}  %MikTeX
\usepackage[cp1251]{inputenc}                            %Win
\usepackage{amsmath,amsfonts,amssymb,amscd,euscript}
\usepackage[russian]{babel}

\usepackage{graphicx}
\usepackage{psfrag}
\makeatletter

\def\titlerus{\thispagestyle{empty} { } \vspace{-5mm} \noindent
\raisebox{-37pt}[\headheight][0pt]{\vbox{ \hbox to \textwidth{\hfil
\scriptsize  ВЕСТНИК \; УДМУРТСКОГО \; УНИВЕРСИТЕТА\hfil }
\vspace{2pt} \hrule \vspace{8pt} \hbox to \textwidth{\series \hfil  \issue}
\vspace{30pt} \hbox{УДК \UDK} }} \vspace{ 30pt plus 6pt }}

\def\titleeng{\vspace{3ex} \hfill Поступила  в редакцию \datereceive \par \vspace{5ex} \par
\noindent \parbox{166mm%130mm
}{\small {\textbf {\textit {\autorseng}}} \par {\bf \articleseng} \par
\vspace{10pt} \par \annotationeng \par \vspace{7pt} \par {\it Keywords}: \keywordseng
\par \vspace{7pt} \par \noindent \small {Mathematical Subject Classifications}: \MSC }
\par \vspace{30pt} \par \small \noindent \contactinformation}

\def\annotationandkeywordsrus{\noindent {\small \annotationrus \par } \vspace{8pt}
\noindent {\small {\it Ключевые слова}: \keywordsrus} \par \vspace{10pt}}

\@addtoreset{equation}{section}
\renewcommand{\section}{\@startsection{section}{1}{0pt}{1.3ex
plus 1ex minus .1ex}{1.3ex plus .1ex}{\bf\,\S\,}}

\renewcommand{\@begintheorem}[2]{\begin{trivlist}
\item[\hspace{\labelsep}{\bf \mbox{~~~}#1\ #2.}]}
\renewcommand{\@opargbegintheorem}[3]{\begin{trivlist}
\item[\hspace{\labelsep}{\bf \mbox{~~~}#1\ #2 {\rm (#3).}}]}
\renewcommand{\@endtheorem}{\end{trivlist}}

\newtheorem{teo}{Теорема}

\newtheorem{pre}{Предложение}

\newcommand{\doc}{\mbox{Д о к а з а т е л ь с т в о}}
\renewcommand{\@evenfoot}{}
\renewcommand{\@oddfoot}{}

\newcommand{\re}{\vspace{-0.5em}}
\renewcommand*{\@biblabel}[1]{#1.\hfill}

\newcommand*{\CSep}{.\ }
\renewcommand{\@makecaption}[2]{%
  \vskip\abovecaptionskip
  \sbox\@tempboxa{{\bf #1\CSep}{#2}}%
  \ifdim \wd\@tempboxa >\hsize
  \begin{center}%
    {\footnotesize{\bf #1\CSep}{#2\par}}%
  \end{center}%
  \else
    \global \@minipagefalse
    \hb@xt@\hsize{\hfil\box\@tempboxa\hfil}%
  \fi
  \vskip\belowcaptionskip%
}

\renewcommand{\@evenhead}{\raisebox{0pt}[\headheight][0pt]{\vbox{\hbox to\textwidth{\thepage \strut \hfil
\text{\autorsrus} \hfil } \hrule \vspace{8pt} \hbox to \textwidth{\series \hfil  \issue}}}}

\renewcommand{\@oddhead}{\raisebox{0pt}[\headheight][0pt]{\vbox{\hbox to\textwidth{  \strut \hfil
\text{\articleshortname} \hfil \thepage} \hrule \vspace{8pt} \hbox to \textwidth{\series \hfil  \issue}}}}

\newcommand{\series}{КОМПЬЮТЕРНЫЕ НАУКИ}

\newcommand{\issue}{2011. Вып.\,2} %%% Год и номер выпуска.

\newcommand{\autorsrus}{И.\,И.~Харламовa, П.\,Е.~Рябов} %%% на русском языке (точку в конце не ставим)
\newcommand{\autorseng}{I.\,I.~Kharlamova, P.\,E.~Ryabov} %%% на английском языке (точку в конце не ставим)

\newcommand{\articleshortname}{Электронный атлас бифуркационных диаграмм }
\newcommand{\articleseng}{The electronic atlas of bifurcation diagrams of the Kowalevski--Yehia gyrostat}

\newcommand{\UDK}{517.938.5+531.38} %%% Проставляет автор!!!
\newcommand{\MSC}{70E17, 70G40} %%% Проставляет автор!!!

\newcommand{\annotationrus}{Рассматривается интегрируемый случай Ковалевской--Яхья в динамике гиростата. Представлен новый подход к классификации бифуркационных диаграмм приведенных систем. Получены конструктивно проверяемые условия существования критических движений на сечении фиксированной постоянной площадей поверхностей, несущих бифуркационную диаграмму трех интегралов полной исходной системы. Случаи, когда эти условия претерпевают качественные перестройки, дают аналитические зависимости между постоянной площадей и величиной гиростатического момента, формирующие разделяющее множество в плоскости двух параметров семейства диаграмм приведенных систем. В результате создана компьютерная система,  удовлетворяющая введенному понятию электронного атласа.
}

\newcommand{\annotationeng}{
The integrable case of Kowalevski--Yehia in the dynamics of a gyrostat is considered. We present the new way to classify the bifurcation diagrams of the reduced systems. We find the efficiently checked existence conditions for the critical motions on the area integral constant sections of the surfaces bearing the 3-diagram of the complete system. The cases when these conditions qualitatively change give the analytical expressions of the dependencies between the area constant and the gyrostatic momentum forming the classifying set for the two-parametric family of the reduced systems diagrams. Finally, we present the computer system, which satisfy the given definition of the electronic atlas.
}

\newcommand{\keywordsrus}{интегрируемые гамильтоновы системы, бифуркационная диаграмма, электронный атлас.}
\newcommand{\keywordseng}{integrable Hamiltonian system, bifurcation diagram, electronic atlas.}

\newcommand{\datereceive}{17.03.11} %%% формат дд.мм.гг

\newcommand{\contactinformation}{Харламова Ирина Ивановна, к.\,ф.-м.\,н.,
доцент, кафедра математического моделирования,
Волгоградская академия госслужбы, 400131,
Россия, г. Волгоград, ул. Гагарина,\,8, E-mail:
irinah@vags.ru\\
Рябов Павел Евгеньевич, к.\,ф.-м.\,н., доцент,
кафедра теории вероятностей и математической
статистики, Финансовый университет при
Правительстве Российской Федерации, 125993,
Россия, г. Москва,  Ленинградский просп., 49,
E-mail: orelryabov@mail.ru}
\newcommand{\la}{{\lambda}}

\newcommand{\A}{{\gamma}}

\newcommand{\ri}{{\rm i}\,}
\newcommand {\mP}{P}
\newcommand {\mD}{\mathcal{D}}

\newcommand {\mla}{P^4_\ell(\la)}
\newcommand {\ml}{P^4_\ell}

\newcommand {\mA}{\Theta}

\newcommand {\bbR}{\mathbb{R}}
\newcommand {\bbI}{\bbR^3(h,k,\ell)}
\newcommand {\sil}{\mathcal{S}_\ell}
\newcommand {\sila}{\mathcal{S}_\ell(\lambda)}

\newcommand {\silaj}{\mathcal{S}_{j}(\la,\ell)}
\newcommand {\sone}{\mathcal{S}_1(\la,\ell)}

\newcommand {\sthr}{\mathcal{S}_3(\la,\ell)}

\newcommand {\pil}{\mathcal{P}_\ell}

\newcommand {\wsi}{\Pi}
\newcommand {\bfa}{\Pi_1}
\newcommand {\crito}{\mathfrak{C}_0}
\newcommand {\criti}{\mathfrak{C}_1}
\newcommand {\crit}{\mathfrak{C}}
\newcommand {\ma}{\mathcal{M}_1}
\newcommand {\bfb}{\Pi_2}
\newcommand {\mb}{\mathcal{M}_2}
\newcommand {\bfc}{\Pi_3}
\newcommand {\mc}{\mathcal{M}_3}

\newcommand {\gan} {\delta_0}
\newcommand {\ltan} {\ell_{\rm tan}}
\newcommand {\Fun} {Z}
\newcommand{\cons}{\mathop{\rm const}\nolimits}
\newcommand{\rk}{\mathop{\rm rank}\nolimits}
\renewcommand{\cosh}{\mathop{\rm ch}\nolimits}
\renewcommand{\sinh}{\mathop{\rm sh}\nolimits}
\newcommand {\vk}{\varkappa}
\newcommand {\ro}{\rho}
\newcommand {\ds}{\displaystyle}
\newcommand{\gs}{\geqslant}
\newcommand{\ls}{\leqslant}
\newcommand{\sgn}{\mathop{\rm sgn}\nolimits}
\newcommand {\mstrut}{\vphantom{\bigl(}}
\newcommand{\bs}{\boldsymbol}

\begin{document}

\titlerus

\begin{flushleft}
{\bf \copyright { \textit { \autorsrus}} \\[2ex]
{ЭЛЕКТРОННЫЙ АТЛАС БИФУРКАЦИОННЫХ ДИАГРАММ ГИРОСТАТА КОВАЛЕВСКОЙ--ЯХЬЯ}%%% ПОЛНОЕ! НАЗВАНИЕ СТАТЬИ
%%% в следующих трех строках Вы проставляете "свои" гранты. Если Ваша работа
%%% не поддержана грантами, эти строки надо закомментировать (лучше стереть)
\footnote{Работа выполнена при финансовой
поддержке РФФИ и АВО (грант 10--01--97001).} }
\end{flushleft}

\annotationandkeywordsrus

%----------------------------------
%\sect{\label{sec1}Постановка задачи}

\begin{flushleft}{\bf{Введение}}\end{flushleft}
% Введение не нумеруется (поэтому команда \sect не применяется)
% с красной строки
%%%%%%%%%%%%%%%%%%%%%%%%%%%%%%%%%%%%%%

Случаем Ковалевской--Яхья называют задачу о движении тяжелого гиростата, главные моменты инерции которого удовлетворяют отношению 2:2:1, центр масс лежит в экваториальной плоскости, а гиростатический момент направлен по оси динамической симметрии. Подходящим выбором осей и единиц измерения уравнения движения приводятся к виду
\begin{equation}\label{eq1_1}
\begin{array}{lll}
2\dot\omega _1   = \omega _2 (\omega _3- \la)  , &
2\dot\omega _2 =  - \omega _1 (\omega _3-\la)  - \alpha _3 , &
\dot\omega _3   = \alpha _2, \\[2mm]
\dot\alpha _1   = \alpha _2 \omega _3  - \alpha _3 \omega _2 , &
\dot\alpha _2   = \alpha _3 \omega _1  - \alpha _1 \omega _3 , &
\dot\alpha_3   = \alpha_1 \omega_2  - \alpha_2 \omega_1,
\end{array}
\end{equation}
где $\la > 0$. Фазовое пространство $\mP^5=\bbR^3_{\omega}{\times}S^2_{\alpha}$ определено в $\bbR^6$ геометрическим интегралом ${\bs \alpha}^2=1$. Частные решения системы \eqref{eq1_1}, формирующие трехмерные подмногообразия, были получены в работах \cite{PVLect, PVMtt71}. Х.М.\,Яхья указал, в дополнение к классическим интегралам энергии и площадей, новый интеграл типа Ковалевской, получив полную инволютивную систему \cite{Yeh1}
\begin{equation}\label{eq1_2}
\begin{array}{l}
H = \omega _1^2  + \omega _2^2 + \ds{\frac{1}{2}}\omega _3^2 -
\alpha _1, \qquad
L =\omega _1 \alpha _1  + \omega _2 \alpha _2  + \frac{1}{2}(\omega _3+\la) \alpha _3,
\\[2mm]
K=(\omega_1^2-\omega^2_2+\alpha_1)^2+(2\omega_1\omega_2+\alpha_2)^2 + 2\la[(\omega_3-\la)(\omega_1^2+\omega^2_2)+2\omega_1 \alpha_3].
\end{array}
\end{equation}

Исследование множества критических точек интегрального отображения
\begin{equation}\label{eq1_3}
J=H{\times}K{\times}L: \mP^5 \to \bbR^3
\end{equation}
начато в \cite{Gash1,Ryab2,Gash3} и завершено в работах \cite{Gash4,Ryab1}. Как оказалось, это множество исчерпывается решениями П.В.\,Харламова. В \cite{Gash4,Ryab1} получены  уравнения бифуркационных поверхностей, т.е. связных поверхностей $\wsi_j$ в $\bbI$, объединение которых $\wsi$ содержит в себе бифуркационную диаграмму $\Sigma$ интегралов $H,K,L$ как собственное подмножество. Здесь и далее постоянные первых интегралов обозначаются строчными буквами, соответствующими обозначениям самих интегралов как функций на фазовом пространстве. В работах \cite{Ryab5, Ryab6, Ryab7} исследовалась эволюция сечений $\pil$, $\sil$ множеств $\wsi$, $\Sigma$ плоскостями $\ell=\cons$.
Эти сечения имеют особое значение, поскольку ограничение системы \eqref{eq1_1} на любое четырехмерное симплектическое многообразие
\begin{equation}\label{eq1_4}
\ml=\{ L = \ell \} \subset \mP^5
\end{equation}
является вполне интегрируемой гамильтоновой системой с двумя степенями свободы (и называется, обычно, {\it приведенной системой}). При этом множество $\sil$, рассматриваемое как подмножество в плоскости $(h,k)$, служит бифуркационной диаграммой интегрального отображения
\begin{equation}\label{eq1_5}
J_\ell=H{\times}K|_{\ml}: \ml \to \bbR^2
\end{equation}
приведенной системы, параметризованной постоянной площадей.

В работе \cite{Gash2} указана топология регулярных интегральных многообразий для точек $(h,k,\ell)$ из связных компонент $\bbR^3 \backslash \Sigma$. Все перечисленные объекты и свойства зависят от одного свободного параметра -- величины гиростатического момента $\la$. В связи с
этим в работе \cite{RyabRCD} исследована зависимость диаграммы $\sila$ от {\it двух} параметров $(\la, \ell)$ и на плоскости этих параметров построено множество, при пересечении точек которого меняется тип $\sila$. Кроме того, в \cite{RyabRCD} для нулевой постоянной площадей построено множество в плоскости $(\la, h)$, классифицирующее типы графов Фоменко \cite{FomGr} на трехмерных изоэнергетических уровнях, вычислены эти графы и соответствующие неоснащенные молекулы. Описание всех графов и молекул, включая случай $\ell \ne 0$, приведено в работе \cite{Gash5}. Совокупность результатов по топологическим инвариантам случая Ковалевской--Яхья подробно изложена в \cite[гл. 9]{GashDis}.
Наиболее полное исследование бифуркационных диаграмм содержится в \cite{RyabDis}, где, в частности, в терминах некоторых вспомогательных параметров решается задача определения  областей существования движений на бифуркационных поверхностях (см. \cite[\S 5.3]{RyabDis}).

Цель настоящей работы -- построить, развивая перечисленные результаты, {\it электронный атлас} бифуркационных диаграмм $\sila$ интегральных отображений приведенных систем.

Под электронным атласом двухпараметрического семейства геометрических объектов мы понимаем комплекс аналитических результатов и компьютерных модулей, обеспечивающий реализацию следующих функций:

1) вывод на монитор классифицирующего (разделяющего) множества в плоскости параметров с возможностью интерактивного выбора точки на плоскости параметров;

2) детализацию окрестности выбранной точки на плоскости параметров c целью более точного позиционирования этой точки относительно разделяющего множества;

3) компьютерную визуализацию для выбранного неразделяющего значения параметров структурно устойчивого типа самого объекта;

4) детализацию произвольной области объекта.

Для бифуркационных диаграмм интегрируемых систем аналитические результаты должны включать условия, гарантирующие, что в разделяющее множество и в бифуркационные диаграммы не включены точки, для которых не существует вещественных решений исходной системы. В математическом плане получение таких условий оказывается более сложной задачей, чем вывод уравнений бифуркационных поверхностей и чисто геометрическое исследование эволюции сечений этих поверхностей, не связанное с движениями в рассматриваемой механической системе. В связи с этим настоящая работа, в основной своей части, излагает аналитические результаты по условиям существования движений, обеспечивающие корректность работы компьютерной программы визуализации.

Примеры атласов {\it изоэнергетических} диаграмм реализованы для волчка типа Ковалевской в двойном поле (случай Реймана--Семенова-Тян-Шанского \cite{ReySem} без гиростатического момента) по параметрам $(h,\gamma)$, где $\gamma$ -- отношение напряженностей силовых полей \cite{Kh361}, и для гиростата Ковалевской--Яхья по параметрам $(h,\la)$ \cite{KhKhSh40}. Здесь возникает следующая проблема. Пусть $\wsi_j$ -- одна из поверхностей, несущих диаграмму $\Sigma(\beta)$, где $\beta$ -- физический параметр задачи. Пусть $p$ -- постоянная интеграла, на фиксированных уровнях которого строятся диаграммы двух оставшихся интегралов. Для изоэнергетических диаграмм $p=h$. При этом для гиростата Ковалевской--Яхья $\beta=\la$, а для волчка в двойном поле $\beta=\gamma$. Для гиростата Ковалевской--Яхья в рассматриваемой здесь постановке необходимо взять $p=\ell, \beta=\la$.
Предположим, что уравнения для поверхностей $\wsi_j \subset \bbR^3$ записаны в параметрическом виде, необходимом для построения сечений $\mathcal{P}_p(\beta) \supsetneq \mathcal{S}_p(\beta)$ --- постоянные двух интегралов выражены в зависимости от $p$, некоторой второй координаты $s$ и параметра $\beta$. Для построения диаграмм с помощью компьютера необходимо иметь алгоритм, позволяющий при заданном значении $\beta$ по любому~$p$ вычислить промежутки {\it фактического} изменения параметра $s$ на множестве критических точек. Для изоэнергетических диаграмм такой алгоритм реализован в \cite{Kh361}, \cite{KhKhSh40} путем указания области диаграммы на бифуркационных поверхностях неравенствами, в которых $h$ выступает в роли параметра. В случае Ковалевской--Яхья эта проблема для диаграмм $\sila$ до сих пор не была решена.

Точку поверхности $\wsi_j$ будем называть {\it допустимой}, если в ее прообразе относительно отображения момента существуют критические точки. Множество допустимых точек на $\wsi_j$ назовем {\it допустимой областью} и обозначим через $\mD_j$. Ясно, что $\mD_j=\Sigma \cap \wsi_j$. Естественно, что допустимые области, как и все другие рассматриваемые объекты, зависят от физического параметра: $\mD_j=\mD_j(\la)$. Там, где это не приведет к недоразумению, явную зависимость от $\la$ указывать не будем.

В настоящей работе получена вся необходимая информация для компьютерной визуализации $(h,k)$-диаграмм гиростата Ковалевской--Яхья:

- уравнения поверхностей $\wsi_j$ (в данном случае таких поверхностей три) представлены в параметрической форме так, что одним из параметров является постоянная площадей $\ell$, в результате чего второй параметр $s$ оказывается параметром вдоль соответствующей бифуркационной кривой $\silaj=\sila \cap \wsi_j$ в составе диаграммы $\sila$;

- для каждой из поверхностей $\wsi_j$ в плоскости ее параметров $(s,\ell)$ указан образ допустимого множества $\mD_j$;

- для критических подсистем (прообразов поверхностей $\wsi_j$ в множестве критических точек отображения момента) построены отвечающие выбранной параметризации бифуркационные  $(s,\ell)$-диаграммы, изучена их эволюция по $\la$;

- предложен универсальный метод построения разделяющего множества, классифицирующего бифуркационные диаграммы систем, индуцированных на уровне некоторого выбранного интеграла, основанный на свойствах бифуркационных диаграмм критических подсистем;

- указанным методом получены уравнения кривых, формирующих разделяющее множество $\mA$ диаграмм $\sila$, при этом для каждой кривой определены граничные условия, вытекающие из условий существования порождающей особой точки на бифуркационных диаграммах критических подсистем;

- для всех пар $(\la, \ell)$ указаны фактические области $\mD_j(\la, \ell)=\mD_j(\la)\cap \{\ell=\cons\}$ изменения параметра $s$ на кривых $\silaj$.

Даны примеры бифуркационных диаграмм, иллюстрирующие различное количество отрезков в составе $\mD_j(\la, \ell)$. Приведен возможный интерфейс компьютерного модуля, решающего задачу построения электронного атласа.

%\vspace{3mm}

%%%%%%%%%%%%%%%%%%%%%%%%%%%%%%%%%%%%%%%%%%
%\sect{\label{sec2}Критическое множество и бифуркационные поверхности}

\begin{flushleft}\noindent{\bf{\S\,1.\,Критическое множество и бифуркационные поверхности}}\end{flushleft}

Пусть $\crit$ --- множество критических точек отображения \eqref{eq1_3}. Это множество стратифицировано рангом интегрального отображения. В силу того, что интеграл $L$ всюду регулярен и, являясь функцией Казимира для вырожденных скобок Пуассона на $\mP^5$, расслаивает $\mP^5$ на фазовые пространства \eqref{eq1_4} приведенных систем, ранг критической точки вычисляется по отношению к отображению \eqref{eq1_5}. Поэтому полагаем
$$
\crit = \crito \cup \criti, \qquad \crit_i=\{x \in \mP^5: \rk J_\ell (x) = i\}.
$$
С другой стороны, в топологическом анализе оказывается важным разбиение критического множества в соответствии с принадлежностью критических значений к той или иной поверхности, из числа несущих бифуркационную диаграмму. Пересечение прообраза такой поверхности при отображении $J$ с критическим множеством называется {\it критической подсистемой} \cite{KhRCD05,KhND}. Опираясь на результаты цитированных выше работ, представим множество $\crit$ в виде объединения трех критических подсистем ---
инвариантных подмножеств $\ma,\mb,\mc$ в фазовом пространстве, являющихся почти всюду гладкими трехмерными подмногообразиями $\mP^5$. Их образы под
действием $J$ содержатся в поверхностях, обозначаемых соответственно через $\bfa,\bfb,\bfc$. Уравнения этих поверхностей запишем в
форме, вытекающей из представления Лакса \cite{ReySem} при анализе особенностей возникающей алгебраической кривой. Вводя параметр $s$ как удвоенный квадрат
спектрального параметра на кривой, получим (подробности см. в работе \cite{KhND} для более общего случая):
\begin{equation}\label{eq2_1}
\begin{array}{l}
\bfa =
\left\{ \displaystyle{h= \frac{\ell^2}{s^2}+\frac{\la^2}{2} + s,}\;
\displaystyle{k=\frac{\ell^4}{s^4} - \frac{2 \ell^2}{s}+1}, \; \ell s \neq 0
\right\} \cup \\[3mm]
\phantom{\bfa = } \quad \cup \left\{k=1, \; \ell=0 \right\} \cup \left\{\displaystyle{k=1+(h-\frac{\la^2}{2})^2},\;  \ell=0 \right\},
\end{array}
\end{equation}
\begin{equation}\label{eq2_2}
\bfb \cup \bfc = \left\{ \displaystyle{h=2 \ell^2 + \frac{1}{2 s} - \frac{\la^2}{2}(1-4s^2),}\;
k=-4 \ell^2 \la^2 + \displaystyle{\frac{1}{4 s^2}} - \displaystyle{\frac{\la^2}{s}(1-\la^2s)(1-4s^2)}
\right\}.
\end{equation}
Здесь $s <0 $ для $\bfb$ и $s > 0 $ для $\bfc$.

Поверхности $\bfb, \bfc$, очевидно, связны. Поверхность $\bfa$, несмотря на представление в
виде объединения множеств, также имеет одну компоненту связности. Действительно, запишем ее
уравнения, выбрав в качестве параметров $h,s$:
\begin{equation}\label{eq2_3}
\displaystyle{\ell^2=(h-\frac{\lambda^2}{2}-s)s^2,} \qquad \displaystyle{k =
1+(h-\frac{\lambda^2}{2}-s)(h-\frac{\lambda^2}{2}-3s).}
\end{equation}
Никаких особенностей по $s$ в этой записи нет. В точках множества $\crit$ величина $s$ может быть представлена как функция фазовых переменных \cite{KhND}. Обозначим ее через
\begin{equation}\label{eq2_4}
    S: \crit \to \bbR.
\end{equation}
В подсистеме $\ma$ при фиксированном $h$ область определения $S$ есть пересечение $\ma$ с компактным четырехмерным многообразием~-- изоэнергетическим уровнем $\{ H = h\} \subset \mP^5$. В частности, функция $S$ ограничена и, как показано в \cite{KhKhSh40}, ее область значений связна. Полагая в \eqref{eq2_3} $s=h-\la^2/2$ и $s=0$, получим уравнения двух прямых, выделенных в представлении \eqref{eq2_1} в отдельные подмножества.

Движения на $\ma$ соответствуют частному случаю интегрируемости, найденному в \cite{PVLect} для произвольного тензора инерции. Предполагая в этом случае выполненными такие же условия на центр масс, тензор инерции и гиростатический момент, как в системе \eqref{eq1_1}, и выбирая на $\bfa$ точку \eqref{eq2_1} с заданными значениями $\ell,s$, запишем уравнения многообразия $\ma$ в следующем виде (свободные параметры $\ell,s,r$):
\begin{equation}\label{eq2_5}
\begin{array}{lll}
  \ds{\omega_1=p_0,} &  \ds{\omega_2=0,} & \ds{\omega_3 = r,}\\
  \ds{\alpha_1=\frac{1}{2} r^2-\frac{\la^2}{2}-s,} &
  \ds{\alpha_2= R(r), }&
  \ds{\alpha_3 = -p_0 (r-\la),}\\
\end{array}
\end{equation}
где
\begin{eqnarray}
& & \ds{R^2=-\frac{1}{4}r^4-(p_0^2-\frac{\la^2}{2}-s)r^2+2 \la p_0^2 r+1-(\frac{\la^2}{2}+s)^2-p_0^2\la^2}, \nonumber\\
& & p_0 = \left\{ \begin{array}{ll}
- \frac{\ell}{s}, & \ell \, s \neq 0\\
0, & \ell=0, \; s \neq 0\\
\in \bbR, &  \ell=0, \; s =0
\end{array} \right. . \label{eq2_6}
\end{eqnarray}
Динамика на $\ma$ задана уравнением $\dot r=R(r)$. Очевиден следующий критерий существования движений для интегральных параметров.

%%%%%%%%%%%%%%%%%%
\begin{pre}\label{predl1}
{\it При заданных $s,\ell$ и величине $p_0$, выбранной в соответствии с \eqref{eq2_6}, вещественные решения \eqref{eq2_5} существуют тогда и только тогда, когда $R^2(r)\gs 0$ для некоторого $r\in \bbR$.}
\end{pre}

Движения на $\mb$ и $\mc$ описываются решением, найденным в работе \cite{PVMtt71}.  При заданных $\ell,s$ введем следующие обозначения
\begin{equation}\label{def23}
\begin{array}{c}
   \vk ^2 = \ell^2+\la^2 s^2, \quad \ro^2=1-\ds{\frac{2 \vk^2}{s}},  \quad
   \Fun^2=\ds{\frac{1}{2}}\left[\bigl(X+\ds{\frac{\la}{\vk}}\bigr)^2+\bigl(\ro Y +\ds{\frac{\ell}{s \vk}}\bigr)^2-1\right],\\
   (X,Y)=\left\{ \begin{array}{ll}
   (\cos \sigma, \sin \sigma), & \ro^2 \gs 0 \\
   (\cosh \sigma, \ri \sinh \sigma), & \ro^2 < 0
   \end{array}\right. .
\end{array}
\end{equation}
Здесь $\ri$ --- мнимая единица, $\sigma$ --- вспомогательная переменная. Многообразия $\mb,\mc$ описываются уравнениями (свободные параметры $\ell,s,\sigma$)
\begin{equation}\label{eq2_7}
\begin{array}{lll}
  \ds{\omega_1=-\frac{\ell} {s}- \vk \ro Y,} &
  \ds{\omega_2=-\ro \sqrt{s}\, \Fun,} &
  \ds{\omega_3 = \la+2 \vk X,}\\[2mm]
  \ds{\alpha_1=\frac{\la s X+\ell \ro Y}{\vk} -2\vk^2 Y^2 ,} &
  \ds{\alpha_2=-2 \vk Y \sqrt{s} \, \Fun, } &
  \ds{\alpha_3 = \frac{\ell X-\la s \ro Y }{\vk},}
\end{array}
\end{equation}
а динамика задана уравнением $\dot \sigma^2  = \sgn(\ro^2)\, s \, \Fun^2$. При этом $s<0$ для $\mb$ и $s>0$ для $\mc$.

Из уравнений \eqref{eq2_7} следует, что $\sgn (\ro^2) =\sgn (Y^2)$ и $\sgn (\ro^2) =\sgn (s \Fun^2)$. Пусть $Y_*=Y$, если $\ro$ вещественно, и $Y_*=\ri Y$, если $\ro$ чисто мнимое. Тогда в плоскости $(X,Y_*)$ кривая $\Gamma_0$, заданная уравнением $X^2+Y^2=1$, необходимость которого вытекает из \eqref{def23}, представляет собой окружность или гиперболу, а кривая $\Gamma_1$, заданная уравнением ${\Fun^2(X,Y) = 0}$, при всех $\ro^2 \ne 0$ есть эллипс. Получаем следующее утверждение.
\begin{pre}\label{predl2}
{\it Для существования вещественных решений \eqref{eq2_7} при заданных $\ell,s$ необходимо и достаточно выполнение следующих условий:

$1)$ при $s<0$ окружность $\Gamma_0$ и эллипс $\Gamma_1$ имеют общую точку$;$

$2)$ при $s>0, \; \ro^2 \gs 0$ окружность $\Gamma_0$ не лежит целиком строго внутри области, ограниченной эллипсом $\Gamma_1$$;$

$3)$ при $s>0, \; \ro^2 < 0$ гипербола $\Gamma_0$ и эллипс $\Gamma_1$ имеют общую точку.}
\end{pre}

Существуют траектории системы \eqref{eq1_1}, принадлежащие одновременно различным критическим подсистемам. А именно, в прообразе точек трансверсального пересечения поверхностей $\wsi_j$ содержится все множество $\crito$, а в прообразе точек касания лежат {\it вырожденные} критические точки из множества $\criti$. Классификация критических точек по рангам и типам дана в работе \cite{RyabUdgu}. Различные формы уравнений множества $\crito$ имеются в работах \cite{Ryab2,Gash4,Ryab1} (подробности можно найти в \cite{RyabDis}). Ниже мы воспользуемся удобной параметризацией $\crito$, предложенной И.Н.\,Гашененко \cite{Gash4}.

%%%%%%%%%%%%%%%%%%%%%%%%%%%%%%
% \sect{\label{sec3}Бифуркационные диаграммы критических подсистем}
\begin{flushleft}
\noindent{\bf{\S\,2.\,Бифуркационные диаграммы критических подсистем}}
\end{flushleft}

Перестройкам интегральных многообразий {\it внутри} критических
подсистем отвечают некоторые особенности объединения поверхностей~\eqref{eq2_1}, \eqref{eq2_2}. Геометрия этих
поверхностей изучалась в работах \cite{Ryab2,Gash4,Ryab1,RyabDis}. Нам понадобятся уточнения и обозначения особых множеств, связанные с условиями
существования критических движений. Имеется три кривые, по которым $\bfa$ пересекается с объединением $\bfb \cup \bfc$ трансверсально. В части, принадлежащей $\Sigma$, обозначим их через $\delta_1,\delta_2,\delta_3$ (необходимые формулы будут даны ниже). Кроме этого, поверхность $\bfa$ касается $\bfc$ по кривой, пересечение которой с $\Sigma$ обозначим через
$\gan$. Допустимые области $\mD_j$, кривые $\delta_i$ и возникающие на этих кривых узловые точки, будучи геометрическими объектами в $\bbI$, получают различное представление в координатах на поверхностях $\wsi_j$. Несмотря на это, для их образов в $(s,\ell)$-плоскостях для наглядности сохраним те же обозначения.

Под бифуркационными диаграммами критических подсистем понимаем бифуркационные диаграммы общего интеграла $L$ и частного интеграла \eqref{eq2_4} на соответствующем множестве $\mathcal{M}_j$. Имея в виду решение вопроса о допустимых областях изменения параметра $s$ при любом заданном~$\ell$, определим, какие диаграммы имеют один и тот же тип. Для бифуркационных диаграмм $\sila$ соответствующее определение впервые дано в \cite{Ryab5} (см. также \cite{RyabRCD}). Однако для диаграмм критических подсистем оно требует уточнения.

Обозначим через $\mD_j(\la,\ell)$ область фактического изменения параметра $s$ при фиксированном $\ell$ на бифуркационной диаграмме.
Для краткости это замкнутое подмножество прямой будем называть {\it допустимым промежутком}. Очевидно, $\mD_j(\la,\ell)$ есть сечение прямой $\ell={\rm const}$ допустимой области $\mD_j(\la)$ в плоскости $(s,\ell)$. В $(h,k)$-плоскости, по соответствующим формулам \eqref{eq2_1}, \eqref{eq2_2}, допустимому промежутку отвечает одномерный сегмент (возможно, состоящий из нескольких дуг) диаграммы $\sila$, при этом в узловые точки этой диаграммы переходят пересечения прямых $\ell={\rm const}$ с кривыми $\delta_i$ ($i=0,1,2,3$), образующие конечное множество. Обозначим это множество через $C_j(\la, \ell)$. Допустимый промежуток с отмеченными на нем точками особого множества $C_j(\la,\ell)$, назовем {\it оснащенным допустимым промежутком} и обозначим через $\mD^*_j(\la,\ell)$. Будем считать две диаграммы критической подсистемы однотипными, если существует гомеоморфизм плоскости $(s,\ell)$, монотонный по $\ell$, переводящий первую диаграмму во вторую и каждый оснащенный допустимый промежуток $\mD^*_j(\la,\ell)$ первой диаграммы в оснащенный допустимый промежуток второй диаграммы. Иначе говоря, рассмотрим множество $\mD_j(\la)$ и его проекцию на ось $O\ell$ вдоль оси $Os$ как расслоение $F(\la)$ со слоем $\mD^*_j(\la,\ell)$ и введем очевидное отношение эквивалентности в слоях (как гомеоморфизм множеств $\mD_j(\la,\ell)$, сохраняющий $C_j(\la,\ell)$). Две диаграммы назовем однотипными, если соответствующие расслоения $F(\la)$ в топологическом смысле изоморфны.

Обозначим
\begin{equation}\label{eq3_1}
    \begin{array}{c}
      \varphi_{\pm}(r)= \psi_{\pm}(r) \sqrt{\pi_{\pm}(r)},\quad
      \psi_{\pm}(r)=\ds{\frac{1}{2}\bigl[  \la (r-\la) \pm D\bigr]}, \quad \pi_{\pm}(r)=\ds{\frac{r}{2}\bigl[ -r \pm \frac{1}{r-\la}D\bigr]},\\[2mm]
      D=\sqrt{\mstrut r^2(r-\la)^2+4} \gs 0.
    \end{array}
\end{equation}

\begin{pre}\label{predl3}
{\it Бифуркационная $(s,\ell)$-диаграмма критической системы $\ma$ состоит из следующих множеств:
\begin{equation}\notag
    \begin{array}{lll}
%      \gaa: \quad \ell=0 , & s \gs -1 - \ds{\frac{\la^2}{2}};\\[2mm]
      A_0:  & s=0, & \ell=0; \\
      \delta_1: & s=\psi_-(r), & \ell=\pm \varphi_-(r), \quad r \in [0,\la);\\
      \delta_2: & s=\psi_+(r), & \ell=\pm \varphi_+(r), \quad r \in (-\infty,0];\\
      \delta_3: & s=\psi_+(r), & \ell=\pm \varphi_+(r), \quad r \in (\la,+\infty).
    \end{array}
\end{equation}
При этом границами допустимой области $\mD_1$ $($области существования движений в системе $\ma)$ служат кривые $\delta_1$ и $\delta_3$, а также ось $s=0$, точки которой, за исключением начала координат, в область $\mD_1$ не входят. В области $\la>0$ перестройки диаграмм происходят при следующих значениях параметра: $\la_*=\sqrt[4]{1/8}, \, \la^*=\sqrt[4]{64/27}, \, \sqrt{2}$.}
\end{pre}

\doc. Бифуркациям решений \eqref{eq2_5} по параметрам $s,\ell$, кроме очевидной особой точки $A_0(0,0)$, отвечают точки $(s,\ell)$, в которых многочлен $R^2(r)$ имеет кратный корень. Соответствующие решения с $r=\cons$ полностью исчерпывают множество $\crito$ неподвижных точек системы \eqref{eq1_1}, а для систем с двумя степенями свободы на $\mla$ точки из $\crito$ служат критическими точками ранга 0. Выбирая кратный корень в качестве параметра, на дискриминантном множестве будем иметь
\begin{equation*}
    \begin{array}{c}
     s=\psi_{\pm}(r), \qquad \ell=\pm \varphi_{\pm}(r).
    \end{array}
\end{equation*}
При этом $p_0^2=\pi_{\pm}(r)$. Очевидно, что $\sgn \pi_{\pm}(r) = \pm \sgn r(r-\la).$
Поэтому из предложения~\ref{predl1} получим образ множества $\crito$ в виде кривых $\delta_i$ с указанным в предложении выбором знаков и областей изменения $r$. Этот результат в такой форме впервые представлен в работе \cite{Gash4}.

В множество особых точек $C_1(\la, \ell)$ на сегменте $\mD_1(\la, \ell)$, в дополнение к пересечениям с кривыми $\delta_i$, необходимо добавить и особую точку $A_0$. Очевидно, перестройки на сегменте $\mD_1(\la, \ell)$ происходят при переходе значения $\ell$ через нуль или через критическое значение функции $L$, ограниченной на множество $\crito$, т.е. через экстремум $\ell$ на объединении кривых $\delta_i$. Такие экстремумы определены уравнением
\begin{equation}\label{eq3_2}
\varphi_{\pm}'(r)=0.
\end{equation}
В свою очередь, перестройки бифуркационных $(s,\ell)$-диаграмм системы $\ma$ происходят либо при наличии в области изменения $r$ кратного корня уравнения \eqref{eq3_2}
\begin{equation}\label{eq3_3}
\varphi_{\pm}'(r)=0, \qquad \varphi_{\pm}''(r)=0,
\end{equation}
либо когда корень \eqref{eq3_2} соответствует попаданию на кривую $\delta_i$ точки $A_0$
\begin{equation}\label{eq3_4}
\varphi_{\pm}'(r)=0, \qquad \psi_{\pm}(r)=0.
\end{equation}
Две этих системы уравнений, после исключения $r$, с точностью до ненулевых множителей сводятся, соответственно, к следующим условиям на параметр $\la$:
\begin{equation*}
\la (\la^4-4)(8\la^4-1)=0, \qquad (\la^4-4)(27\la^4-64)=0.
\end{equation*}
Таким образом, особыми значениями $\la$ при классификации $(s,\ell)$-диаграмм системы $\ma$ являются $\la_*=\sqrt[4]{1/8}, \, \la^*=\sqrt[4]{64/27}, \, \sqrt{2}$. Нулевое значение $\la$ как отвечающее особому случаю~--- задаче Ковалевской, здесь не рассматривается.
\hfill $\square$

\begin{figure}[ht]
\centering
\includegraphics[width=150mm, keepaspectratio]{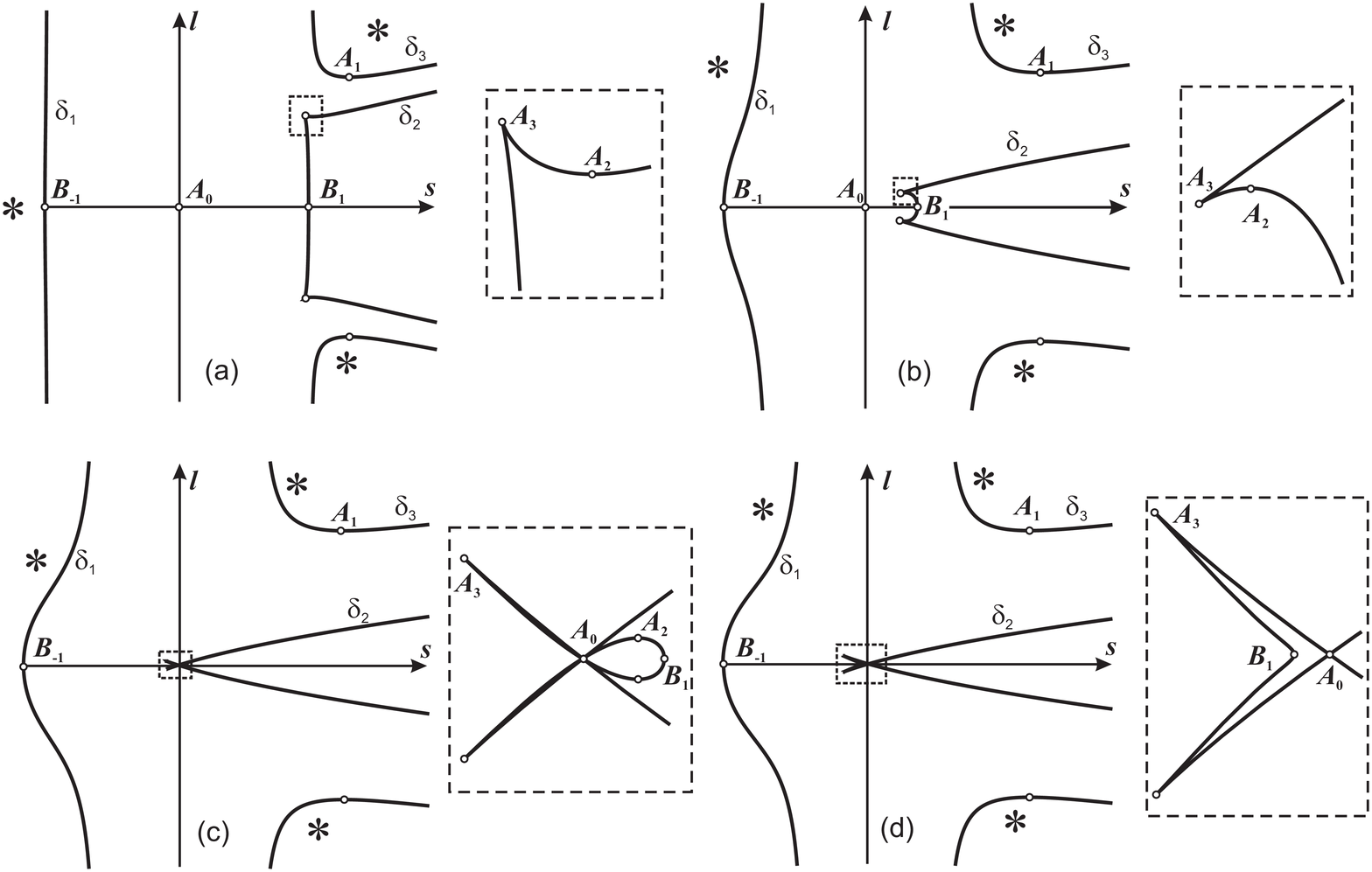}
\caption{Бифуркационные диаграммы на $\ma $.}\label{KhRy_fig1}
\end{figure}

Диаграммы для неразделяющих значений параметра показаны на рис.~\ref{KhRy_fig1}: $(a)$\,${0< \la < \la_*;}$ $(b)$\,${\la_* < \la < \la^*;}$ $(c)$\,$\la^*< \la < \sqrt{2};$ $(d)$\,$\la > \sqrt{2}.$ В прямоугольнике слева от каждой диаграммы показана детализация области, в которой, собственно, и происходит перестройка. Здесь и далее на аналогичных рисунках звездочкой отмечены связные компоненты дополнения диаграммы, не входящие в допустимую область.

Перейдем к рассмотрению систем $\mb,\mc$.  Из \eqref{def23}, \eqref{eq2_7} следует, что бифуркации решений по $\ell,s$ происходят либо в случае, когда $\rho^2=0$, либо при наличии касания кривых второго порядка $\Gamma_0$ и $\Gamma_1$. Обозначая в точке касания $\omega_3=r$ (здесь это константа), получим выражения
\begin{equation}\label{eq3_5}
    \begin{array}{c}
      \ell =\pm \varphi_{\pm}(r),\quad
      s=\theta_{\pm}(r)=\ds{\frac{r-\la}{4\la}\bigl[r(r-\la) \mp D\bigr]},
    \end{array}
\end{equation}
где $D, \varphi_{\pm}$ определены в \eqref{eq3_1}. Это совпадение не случайно, поскольку найденные значения отвечают точкам трансверсального пересечения поверхности $\bfa$
с $\bfb, \bfc$, а в фазовом пространстве --- одним и тем же точкам множества $\crito$. Выражения для $s$ на $\ma$ и на $\mb \cup \mc$ различны, так как они определяют одну из компонент нормали к соответствующей поверхности в общей точке. Из \eqref{eq3_5} с учетом области изменения $r$ следует, что $\delta_1 \cup \delta_3 \subset \bfb$, $\delta_2 \subset \bfc$.

Значения интегральных параметров, отвечающие случаю $\rho^2=0$, формируют введенную выше кривую $\gan$ касания поверхностей $\bfa, \bfc$. На обеих поверхностях она имеет одно и то же уравнение
\begin{equation}\label{eq3_6}
    \ell^2- \ltan^2(s)=0,  \qquad    \ltan = \sqrt{\ds{\frac{s}{2}(1-2\la^2s)}}.
\end{equation}
Включение всех таких значений в бифуркационные обусловлено тем, что при переходе через $\rho=0$ принципиально меняется аналитическая структура решения \eqref{def23}, \eqref{eq2_7} и условия его вещественности. Как следует из результатов \cite{KhND}, кривая $\gan$ также отвечает за вырождение индуцированной симплектической структуры на соответствующем четырехмерном критическом подмногообразии фазового пространства $TSO(3)$ нередуцированной системы, а в \cite{RyabUdgu} показано, что критические точки в прообразе $\gan$ по отношению к системе с двумя степенями свободы на $\mla$, хотя и имеют ранг 1, но являются вырожденными.

\begin{pre}\label{predl4}
{\it Бифуркационная $(s,\ell)$-диаграмма критической системы $\mb$ состоит из следующих множеств:
\begin{equation*}
    \begin{array}{ll}
      \delta_1: \quad s=\theta_-(r), & \ell =\pm \varphi_-(r), \quad r \in [0,\la);\\
      \delta_3: \quad s=\theta_+(r), & \ell =\pm \varphi_+(r), \quad r \in (\la,+\infty).
    \end{array}
\end{equation*}
Внешней границей области существования движений служит $\delta_1$.}
\end{pre}
\doc. При $s<0$ обращение величины $\rho^2$ в ноль невозможно. Анализируя условия предложения~\ref{predl2} в областях, на которые полуплоскость $s<0$ делится кривыми \eqref{eq3_5}, приходим к сформулированному утверждению. \hfill $\square$

\begin{figure}[ht]
\centering
\includegraphics[width=0.2\linewidth, keepaspectratio]{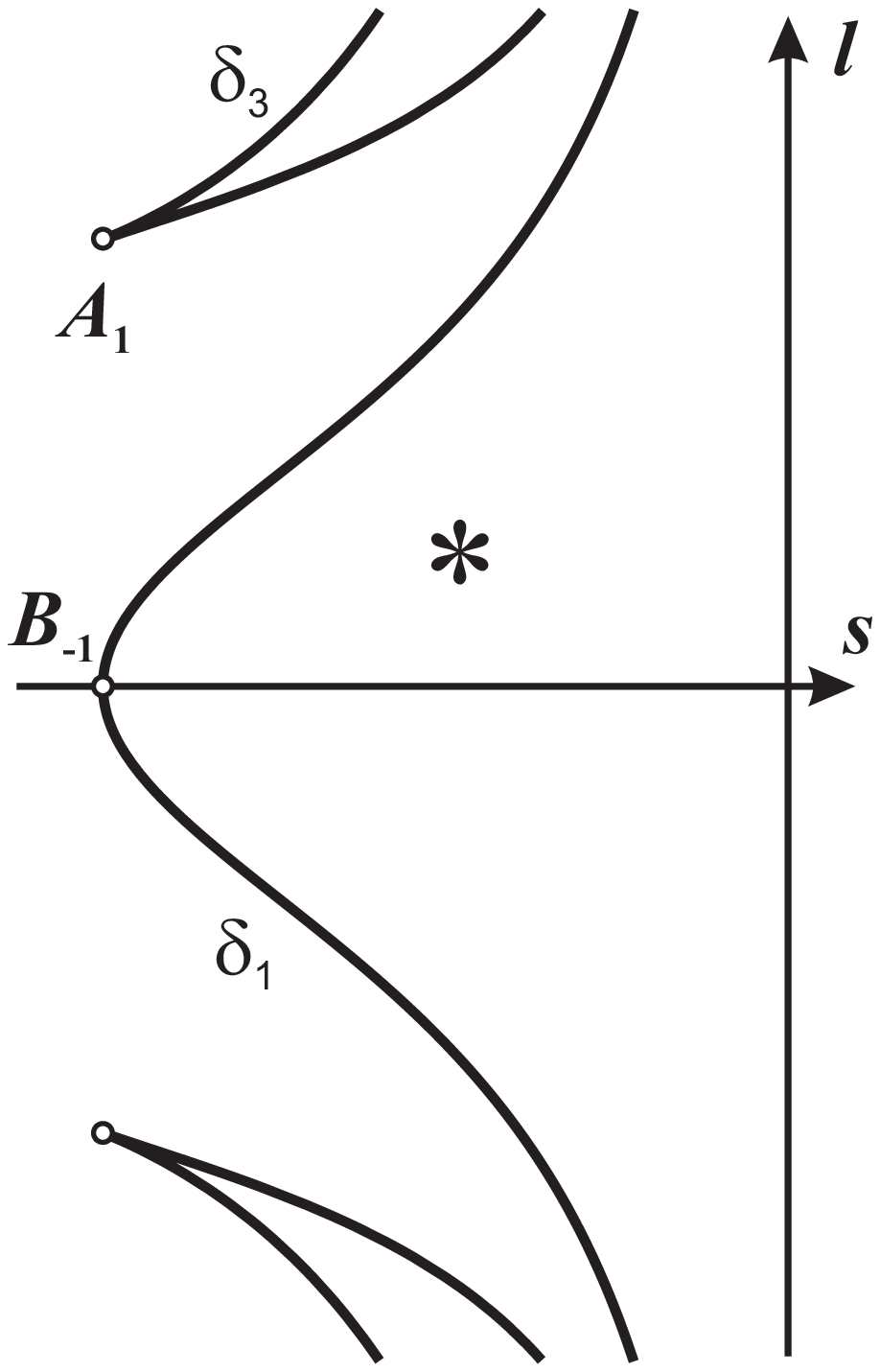}
\caption{Бифуркационная диаграмма на $\mb $.}\label{KhRy_fig2}
\end{figure}

Диаграмма системы $\mb$ показана на рис.~\ref{KhRy_fig2}. Никаких качественных перестроек в ней при $\la \neq 0$ не происходит.

\begin{figure}[ht]
\centering
\includegraphics[width=150mm, keepaspectratio]{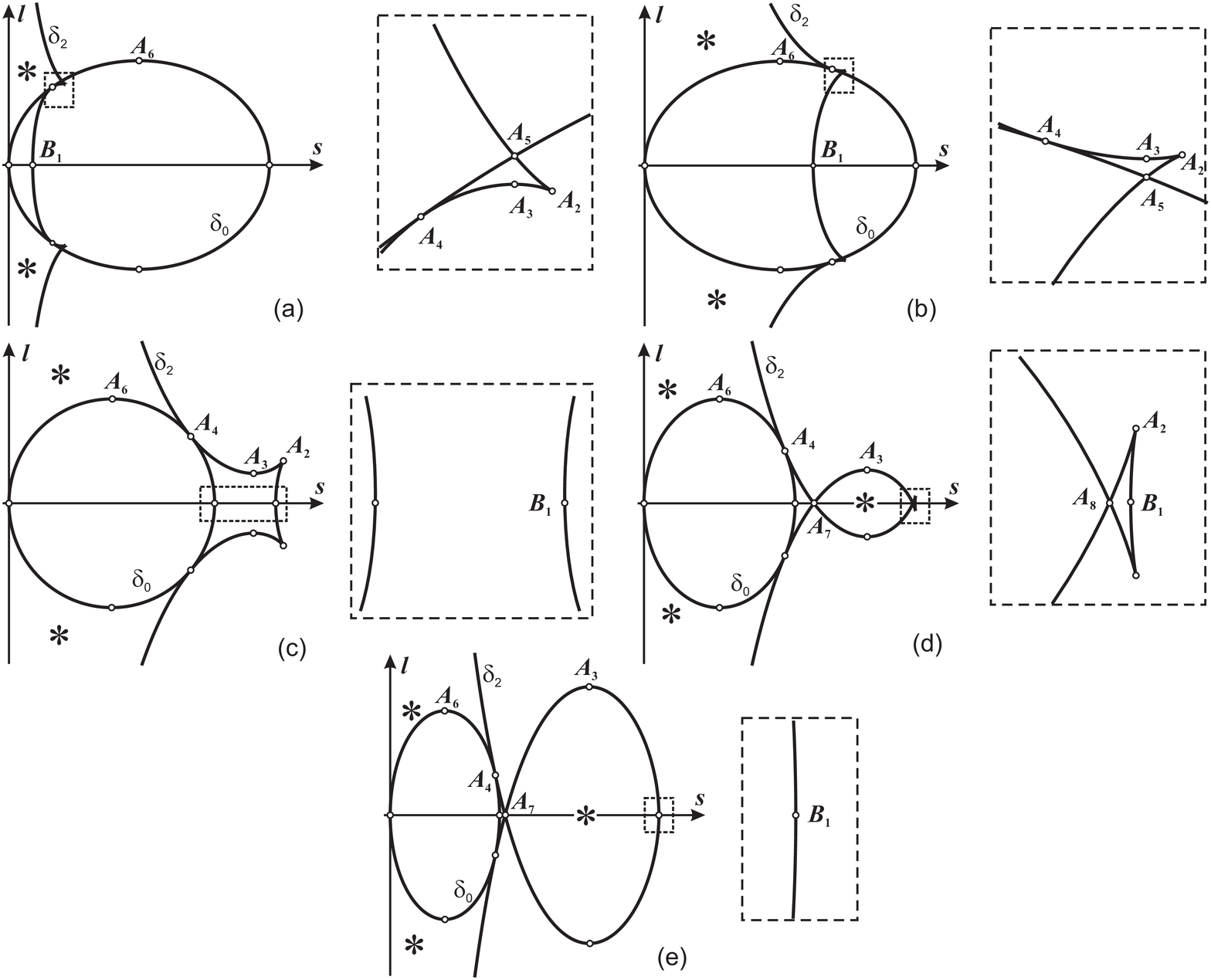}
\caption{Бифуркационные диаграммы на $\mc$.}\label{KhRy_fig3}
\end{figure}

\begin{pre}\label{predl5}
{\it Бифуркационная $(s,\ell)$-диаграмма критической системы $\mc$ состоит из следующих множеств:
\begin{equation*}
    \begin{array}{lll}
      \gan: & \ell=\pm \ltan(s), & 0< s \ls \ds{\frac{1}{2\la^2}};\\[2mm]
      \delta_2: &  s=\theta_+(r), & \ell=\pm \varphi_+(r),  \quad r \in (-\infty,0].
    \end{array}
\end{equation*}
В дополнении к диаграмме имеются следующие области, в которых не существует критических движений: при всех $\la$ --- область, прилегающая к оси $s=0$ и ограниченная ветвями кривых $\delta_0, \delta_2$, при $\la>\la^*$ --- в области, ограниченной кривой $\delta_2$ между двумя ее точками пересечения с осью $\ell=0$ при $r \neq 0$.

Перестройки типов диаграмм в области $\la>0$ происходят при следующих значениях параметра: $\la_*, \, 1,\,\la^*, \, \sqrt{2}$.}
\end{pre}

\doc. В системе $\mc$ величина $\ro^2$, определенная в \eqref{def23}, уже может быть и отрицательной, что влияет на условия существования движений согласно предложению~\ref{predl2}. Отметим равенство, следующее из определений входящих в него величин:
\begin{equation} \label{eq3_7}
\begin{array}{c}
    \ro^2 = \ds{\frac{2}{s}(\ltan^2-\ell^2)}.
\end{array}
\end{equation}
Ограничения на $s$ следуют из \eqref{eq3_6}. Условия существования критических движений проверяются в соответствии с предложением~\ref{predl2}.

Перестройки диаграммы происходят в отмеченных выше случаях \eqref{eq3_3}, \eqref{eq3_4}. В этих уравнениях следует взять верхний знак и $r\in(-\infty,0]$. В дополнение имеем перестройку в множестве особых точек при условиях
\begin{eqnarray}
\varphi_{+}'(r)=0, & \ro^2=0; & \label{eq3_8}\\
r=0 , & \ro^2=0. & \label{eq3_9}
\end{eqnarray}
На кривой $\delta_2$ имеем
$$
\ro^2=\frac{(r+\la)^2}{(r-\la)^2} \frac{ (r-\la)(r-2\la) - D}{ r(r-\la) - D}.
$$
Исключение $r$ из \eqref{eq3_8} приводит к уже имеющемуся случаю $8\la^4-1=0$, а условие \eqref{eq3_9} дает ${\la^2-1=0}$. \hfill $\square$

Диаграммы для неразделяющих значений параметра показаны на рис.~\ref{KhRy_fig3}: $(a)$\,${0< \la < \la_*;}$ $(b)$\,${\la_*< \la < 1;}$ $(c)$\,${1<\la<\la^*;}$ $(d)$\,${\la^*< \la < \sqrt{2};}$ $(e)$\,$\la > \sqrt{2}.$ Как и на рис.~\ref{KhRy_fig1}, слева от каждой диаграммы показана детализация области, в которой происходит перестройка, звездочкой отмечены связные компоненты дополнения диаграммы, не входящие в допустимую область.

%\sect{\label{sec4}Разделяющее множество}
\begin{flushleft}
\noindent{\bf{\S\,3.\,Разделяющее множество}}
\end{flushleft}
Напомним, что оснащенным допустимым промежутком $\mD^*_j(\la,\ell)$ называется сечение образа допустимой области $\mD_j(\la)$ прямой $\{\ell=\cons\}$ с отмеченными на нем точками бифуркационной диаграммы критической подсистемы $\mathcal{M}_j$. Каждой бифуркационной диаграмме $\sila$ интегралов \eqref{eq1_2} соответствует три таких промежутка по количеству критических подсистем. Будем говорить, что две диаграммы, определенные различными точками $P_1, P_2$ плоскости $(\la,\ell)$, однотипны, если существует непрерывный путь из $P_1$ в $P_2$ и однопараметрическое семейство гомеоморфизмов $g_\tau$ плоскости $(h,k)$, переводящих вдоль этого пути диаграмму в диаграмму, не изменяя топологии трех оснащенных допустимых промежутков $\mD^*_j(\la,\ell)$. Такое определение, уточняющее определение, данное в \cite{RyabRCD}, позволяет строить разделяющие множества для классификации бифуркационных диаграмм интегрируемых систем {\it универсальным способом}, исходя из свойств бифуркационных диаграмм критических подсистем, и может быть успешно использовано в ряде задач с аналогичной постановкой \cite{Kh361,Ryab37}.

Пусть $\mA$ --- подмножество в $(\la,\ell)$-плоскости, разделяющее различные типы диаграмм $\sila$. Структура разделяющего множества впервые найдена в работах \cite{Ryab5,Ryab7} (см. также \cite{RyabRCD}).
Так как изначально принято, что $\la>0$, и в силу очевидной симметрии можно считать $\ell \gs 0$, последующие результаты формулируются для первого квадранта плоскости $(\la,\ell)$.
\begin{teo}\label{teor-1}
{\it Множество $\mA$ состоит из следующих кривых:}
\begin{equation*}
\begin{array}{ll}
\A_0: & \ell = 0; \\
\A_{1,2}: & \left\{ \begin{array}{l}
\ds{\la=\la_{1,2}(u)=\left|\ds{\frac{3u^2-4}{2 u^{3/2}}}\right| }\\
\ds{\ell=\ell_{1,2}(u)=\frac{1}{4}\left[\frac{4-u^2}{u}\right]^{3/2}}
\end{array} \right., \quad u \in \left\{\begin{array}{ll}\, (0,2/\sqrt{3}] & \textrm{для\ } \A_1 \\
    \, [2/\sqrt{3},2] & \textrm{для\ } \A_2 \end{array} \right. ;\\
\A_3: & \ell=\ell_3(\la)= \ds{ \frac{|\sqrt{4 + \la^{4/3}}-2\la^{2/3}|}{\sqrt{2}(\sqrt{4 + \la^{4/3}}-\la^{2/3})^{1/2}}},\quad \la \gs 0;\\
\A_4:  & \ell=\ell_4(\la)= \ds{\frac{(\sqrt{1+\la^4}-\la^2)^{3/2}}{\sqrt{2}}}, \quad  \la \gs 0;\\
\A_5:  & \ell=\ell_5(\la)= \ds{\frac{\sqrt{1-\la^{4/3}}}{2\la^{1/3}}}, \quad 0\ls \la \ls 1.\\
\A_6:  & \ell=\ell_6(\la)= \ds{\frac{1}{4\la}}, \quad \la \gs 0.
\end{array}
\end{equation*}
\end{teo}

\doc. Как следует из введенного понятия однотипных диаграмм, разделяющие кривые выражают зависимость между $\la$ и $\ell$ в особых точках бифуркационных диаграмм критических подсистем -- экстремумах $\ell$ на кривых в составе этих диаграмм. Все такие точки отмечены на рис.~\ref{KhRy_fig1}--\ref{KhRy_fig3}.
Очевидно, что при всех $\la$ оснащенные промежутки $\mD^*_j(\la,0)$ не гомеоморфны близлежащим, например, в силу наличия в бифуркационных диаграммах критических подсистем особой точки $A_0$ --- начала координат плоскости $(s,\ell)$. Это дает луч $\A_0$.

Точки $B_{-1}, B_1$ с координатами $(\mp 1-\la^2/2,0)$ при любом $\la$ отвечают глобальным критическим значениям энергии на всем фазовом пространстве, но, не являясь экстремумами функции $L$ (даже условными экстремумами на каких-либо из рассматриваемых подмногообразий), не порождают перестроек допустимых множеств.

Рассмотрим диаграмму критической подсистемы $\ma$. Уравнение \eqref{eq3_2} для нахождения экстремумов $\ell$ на кривых $\delta_i$ ($i=1,2,3$), после сокращения на отличный от нуля множитель ${r(r-\la) \mp D}$,  распадается на два
\begin{eqnarray}
& &(2r-\la)(r-\la) \mp D=0, \label{eq4_1}\\
& &r(2r-\la)(r-\la) \pm \la D=0. \label{eq4_2}
\end{eqnarray}
Из \eqref{eq4_1} получим уравнение
\begin{equation}\label{eq4_3}
    (r-\la)^3(3r-\la)-4=0,
\end{equation}
которое имеет ровно два корня $r_* \in (-\infty, \la/3),  r^* \in (\la,+\infty)$. Тогда $r^*$ --- единственный экстремум $\ell$ на $\delta_3$ (точка $A_1$). При $\la \ls \sqrt{2}$ имеем $r_* \ls 0$, т.е. это --- экстремум $\ell$ на $\delta_2$ (точка~$A_2$). Если $\la > \sqrt{2}$, то $r_*$ лежит в интервале $(0,\la/3)$ и $(2r-\la)(r-\la)>0$. Интервал соответствует кривой $\delta_1$, но тогда при выборе нижнего знака уравнение \eqref{eq4_1} несовместно. Итак, точка $A_2$ исчезает при переходе $\la$ через $\sqrt{2}$.
Явной функции $\ell(\la)$ в точках $A_{1,2}$ построить не удается. Поступим следующим образом. Положим
в уравнении \eqref{eq4_3}
\begin{equation}\label{eq4_4}
    x=\la - r.
\end{equation}
Из \eqref{eq4_3} выразим $\la(x)$, а из \eqref{eq4_4} найдем $r(x)$ и подставим в выражение для $\ell$ на кривых $\delta_2,\delta_3$. Получим
\begin{equation*}
    \la = \ds{\frac{3x^4-4}{2x^3}}, \qquad r = \ds{\frac{x^4-4}{2x^3}}, \qquad \ell  = \pm \frac{(4-x^4)\sqrt{4-x^4}}{4x^3},
\end{equation*}
где $x\in [-\sqrt[4]{4/3}\, ,0)$ для $A_1$ и $x\in [\sqrt[4]{4/3} \, , \sqrt{2}\,]$ для $A_2$. Поскольку $\la >0$, можно записать единую параметризацию зависимости $\ell$ и $\la$, положив $u=x^2$. Получим кривые $\A_{1,2}$.

Левая часть уравнения \eqref{eq4_2} с точностью до ненулевого множителя совпадает с $\psi_{\pm}'(r)$, поэтому решения \eqref{eq4_2} определяют точки возврата на $(s,\ell)$-диаграмме системы $\ma$. Как следствие \eqref{eq4_2} получим
\begin{equation}\label{eq4_5}
    r^3(r-\la)^3=\la^2.
\end{equation}
Это уравнение имеет ровно два корня, по одному в интервалах $(-\infty,0)$, $(\la,+\infty)$. Поэтому в уравнении \eqref{eq4_2} нужно брать верхний знак (см. предложение~\ref{predl3}). Значит, ${r(2r-\la)(r-\la)<0}$ и допустимым является лишь отрицательный корень \eqref{eq4_5}, дающий точку $A_3$ на~$\delta_2$. Из~\eqref{eq4_5} и уравнения кривой $\delta_2$ получим зависимость на кривой $\A_3$, или, в удобной параметрической форме,
\begin{equation*}
\A_3: \quad \la = \ds{\left(\frac{1}{z}-z
\right)^{3/2}}, \qquad \ell= \ds {\frac{|3z^2-1|}{2 z^{3/2}}}, \qquad z \in (0,1].
\end{equation*}

Диаграмма на $\mb$ новых разделяющих зависимостей между $\ell$ и $\la$ в состав $\mA$ не добавляет, поскольку единственная узловая точка $A_1$ уже учтена в системе $\ma$.
Отметим интересную двойственность, типичную для случая, когда ребро возврата одной бифуркационной поверхности попадает на другую. Возникающий след в параметрах одной поверхности дает экстремум на гладкой кривой, а в параметрах другой порождает точку возврата. Здесь это видно для точки $A_1$ в первой и второй системах.

Обратимся к системе $\mc$. Особенности $A_2,A_3$ на кривой $\delta_2$ уже учтены. Здесь снова проявляется отмеченная двойственность: точка $A_2$, бывшая в диаграмме системы $\ma$ простым экстремумом $\ell$, становится точкой возврата для диаграммы на $\mc$. С точкой $A_3$ происходит обратное явление.

Кривые $\delta_0$ и $\delta_2$ имеют точку касания $A_4$ при всех $\la$ и точку трансверсального пересечения $A_5$ при $\la<1$. Вычисляя значения $\ell$, получим уравнения кривых $\A_4, \A_5$. Отметим также точку $A_6$ максимума $\ell$ на кривой $\delta_0$ при $s=1/(4\la^2)$. Из \eqref{eq3_7} в этой точке получим кривую $\A_6$.
Других случаев перестройки оснащенных допустимых промежутков при изменении $\ell$ нет. Теорема доказана. \hfill $\square$

\clearpage

%\sect{\label{sec5}Классификация допустимых промежутков}
\begin{flushleft}
\noindent{\bf{\S\,4.\,Классификация допустимых промежутков}}
\end{flushleft}
Следующая теорема об устройстве допустимых множеств на поверхностях $\bfa,\bfb,\bfc$ сформулирована таким образом, что в ней для любого $\ell$ явно определены допустимые промежутки по $s$. Это позволяет непосредственно реализовать компьютерный алгоритм отсечения лишних отрезков на $\ell$-сечениях поверхностей, т.е. построить любую диаграмму $\sila$.

Введем следующую договоренность. При фиксированном $\la$ проведем сечение образа $\delta_i$ в $(s,\ell)$-плоскости, отвечающей поверхности $\wsi_j$, на заданном уровне $\ell$. Если на рассматриваемой ветви кривой точка пересечения единственна, то ее \mbox{$s$-}координату обозначим через $\eta_{ij}(\ell)$. Если же таких точек две, то их \mbox{$s$-}координаты обозначим через $\eta_{ij}^-(\ell)< \eta_{ij}^+(\ell)$. Обозначения для случаев, когда имеется большее количество точек пересечения, нам не понадобятся. Зависимость от $\la$ подразумевается. Зависимость от $\ell$ для краткости в формулировке теоремы опускаем.

\begin{teo}\label{teor-2}
{\it Допустимая область на поверхности $\bfa$ описывается следующим образом:}
\begin{equation*}
    \begin{array}{lll}
      \ell=0   & \Rightarrow & s \in [-1-\la^2/2, +\infty);\\
      0< |\ell| \ls \ell_3 & \Rightarrow & s \in [\eta_{11},0)\cup(0,+\infty);\\
      |\ell| > \ell_3  & \Rightarrow & s \in [\eta_{11}, 0)\cup (0,\eta^-_{31}] \cup [\eta^+_{31}, +\infty].
    \end{array}
\end{equation*}

{\it Допустимая область на поверхности $\bfb$ описывается неравенством}
\begin{equation*}
    s \ls \eta_{12}.
\end{equation*}

{\it Допустимая область на поверхности $\bfc$ описывается следующими неравенствами:

{\noindent при $0< \la \ls \la_*$}
\begin{equation*}
    \begin{array}{lll}
      \ell \in [0,\ell_7] & \Rightarrow & s \in [\zeta^-,+\infty);\\
      \ell \in (\ell_7,+\infty) & \Rightarrow & s \in [\eta_{23},+\infty);
    \end{array}
\end{equation*}
при $\la_* < \la \ls  \la^*$
\begin{equation*}
    \begin{array}{lll}
      \ell \in [0,\ell_6] & \Rightarrow & s \in [\zeta^-,+\infty);\\
      \ell \in (\ell_6,\ell_8) & \Rightarrow & s \in [\zeta^-,\zeta^+]\cup [\eta_{23},+\infty);\\
      \ell \in [\ell_8,+\infty) & \Rightarrow & s \in [\eta_{23},+\infty);
    \end{array}
\end{equation*}
при $\la^* < \la \ls \la_0$
\begin{equation*}
    \begin{array}{lll}
      \ell \in [0,\ell_5) & \Rightarrow & s \in [\zeta^-,\eta_{23}^-]\cup [\eta_{23}^+,+\infty);\\
      \ell \in [\ell_5,\ell_6] & \Rightarrow & s \in [\zeta^-,+\infty);\\
      \ell \in (\ell_6,\ell_8) & \Rightarrow & s \in [\zeta^-,\zeta^+]\cup [\eta_{23},+\infty);\\
      \ell \in [\ell_8,+\infty) & \Rightarrow & s \in [\eta_{23},+\infty);
    \end{array}
\end{equation*}
при $  \la_0 < \la \ls \la^0$
\begin{equation*}
    \begin{array}{lll}
      \ell \in [0,\ell_6] & \Rightarrow & s \in [\zeta^-,\eta_{23}^-]\cup [\eta_{23}^+,+\infty);\\
      \ell \in (\ell_6,\ell_5) & \Rightarrow & s \in [\zeta^-,\zeta^+] \cup [\eta_{23},\eta_{23}^-]\cup[\eta_{23}^+,+\infty);\\
      \ell \in [\ell_5,\ell_8) & \Rightarrow & s \in [\zeta^-,\zeta^+]\cup [\eta_{23},+\infty);\\
      \ell \in [\ell_8,+\infty) & \Rightarrow & s \in [\eta_{23},+\infty);
    \end{array}
\end{equation*}
при $  \la > \la^0$}
\begin{equation*}
    \begin{array}{lll}
      \ell \in [0,\ell_6] & \Rightarrow & s \in [\zeta^-,\eta_{23}^-]\cup [\eta_{23}^+,+\infty);\\
      \ell \in (\ell_6,\ell_8) & \Rightarrow & s \in [\zeta^-,\zeta^+] \cup [\eta_{23},\eta_{23}^-]\cup[\eta_{23}^+,+\infty);\\
      \ell \in [\ell_8,\ell_5) & \Rightarrow & s \in [\eta_{23},\eta_{23}^-]\cup[\eta_{23}^+,+\infty);\\
      \ell \in [\ell_5,+\infty) & \Rightarrow & s \in [\eta_{23},+\infty).
    \end{array}
\end{equation*}
\end{teo}

\doc. Теорема следует из найденной выше структуры бифуркационных диаграмм критических подсистем. Приведем лишь некоторые пояснения. При $0 \ls \ell \ls \ell_8$ обозначим значения $s$ в точках $\delta_0$ с ординатой $\ell$ так:
\begin{equation*}
    s = \zeta^{\pm}(\ell) = \ds{\frac{1 \pm \sqrt{1-16\la^2\ell^2}}{4\la^2}}.
\end{equation*}

На кривой $\delta_1$ зависимость $\ell(r)$ монотонна, поэтому для любого ${\ell \gs 0}$ существует единственное решение
$r_- = r_-(\la, \ell) \in [0,\la)$ уравнения $\varphi_-(r)=\ell$. Согласно принятым обозначениям, зависимость $s(\ell)$ на $\delta_1$ в подсистемах $\ma,\mb$ определяется функциями $\eta_{11}(\ell)=\psi_-(r_-)$, $\eta_{12}(\ell)=\theta_-(r_-)$.
При $\ell>\ell_3$ уравнение $\varphi_+(r)=\ell$ имеет ровно два вещественных корня в области $r\in (\la,+\infty)$.
Для подсистемы $\ma$ их подстановка в функцию $\psi_+$ дает зависимости
$\eta^-_{31}(\ell)$ и $\eta^+_{31}(\ell)$ на монотонных участках $\delta_3$. Очевидно, $\eta_{11} < \eta^-_{31} < \eta^+_{31}$.
В подсистеме $\mb$ кривая $\delta_3$ на допустимый промежуток не влияет (влияние на оснащенный допустимый промежуток уже учтено).
Подчеркнем, что $\eta_{11}, \eta_{12}, \eta^\pm_{31}$ --- однозначно определенные и эффективно вычисляемые функции от переменной $\ell$ и параметра $\la$.

Для описания допустимой области $\mD_3(\la)$ и ее сечений $\mD_3(\la,\ell)$ уточним систему обозначений для решений на кривой $\delta_2$ системы уравнений
\begin{equation}\label{eq5_1}
    \ell = |\varphi_+(r)|, \qquad s=\theta_+(r) \qquad (r \in (-\infty,0]).
\end{equation}
Снова рассматриваем лишь положительные значения $\la$ и $\ell$. Экстремумы $\ell$ здесь --- найденные ранее значения $\ell_2,\ell_3$ (последнее достигается на $\delta_2$ при $\la \ls \sqrt{2}$). Имеем $\ell_3 > \ell_2$ при $\la < \la_*$ и $\ell_3 < \ell_2$ при $\la_* < \la < \la^*$. При $\la > \la^*$ значение $\varphi_+(r)$ в точке экстремума становится отрицательным, поэтому $\ell_3$ из минимума $\varphi_+$ превращается в максимум $(-\varphi_+)$ (см. случаи $d-e$ на рис.~\ref{KhRy_fig3}).

Пусть $\ell > {\rm max} (\ell_2,\ell_3)$. Тогда решение \eqref{eq5_1} относительно $s$ единственно (монотонно убывающая ветвь кривой $\delta_2$ от $\ell=+\infty$) и обозначается, по сказанному, через $s=\eta_{23}(\ell)$. Пусть $\la > \la^*$ и $0<\ell<\ell_3$. В этом случае нужно рассмотреть систему
\begin{equation*}
    \ell = -\varphi_+(r), \qquad s=\theta_+(r) \qquad (r \in (-\infty,0]),
\end{equation*}
имеющую ровно два решения, и теперь эти значения $s$ обозначим через $\eta_{23}^{-}(\ell) < \eta_{23}^{+}(\ell)$. Предельные значения при $\ell \to 0$ дают, соответственно, точки $A_7, A_8$ при $\la<\sqrt{2}$ (рис.~\ref{KhRy_fig3},$d$) и $A_7, B_1$ при $\la > \sqrt{2}$ (рис.~\ref{KhRy_fig3},$e$).
Отметим еще, что ординаты точек $A_3,A_4$ совпадают при особом значении $\la=\la_*$ (и тогда вдобавок $A_3=A_2$), а также и при некотором $\la = \la_0 \approx 1.383 \in (\la^*,\sqrt{2})$, так что на этом интервале изменения $\la$ имеем $\ell_4> \ell_3$ при $\la < \la_0$ и $\ell_3> \ell_4$ при $\la > \la_0$. Ординаты точек $A_3,A_6$ совпадают при некотором значении $\la = \la^0 \approx 1.515 > \sqrt{2}$, так что $\ell_3 < \ell_6$ при $\la < \la^0$ и  $\ell_3 > \ell_6$ при $\la > \la^0$.
Теорема доказана. \hfill $\square$

\begin{figure}[ht]
\centering
\includegraphics[width=160mm, keepaspectratio]{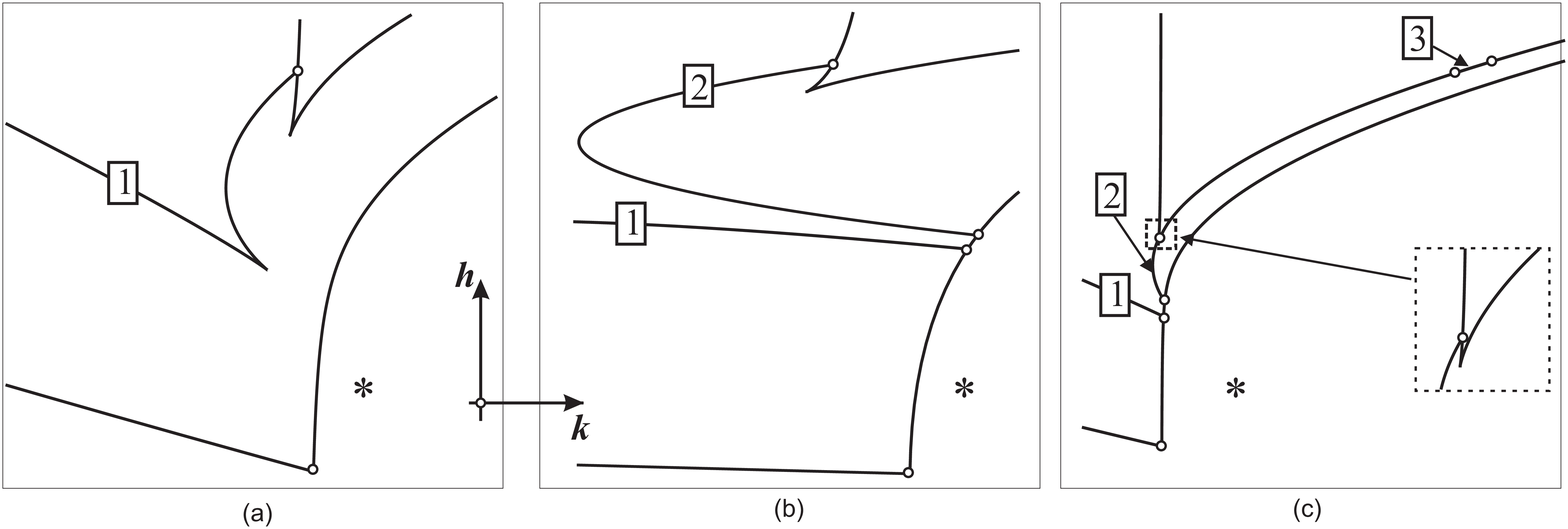}
\caption{Диаграммы с лакунами в третьей подсистеме.}\label{KhRy_fig4}
\end{figure}

\begin{figure}[ht]
\centering
\includegraphics[width=160mm, keepaspectratio]{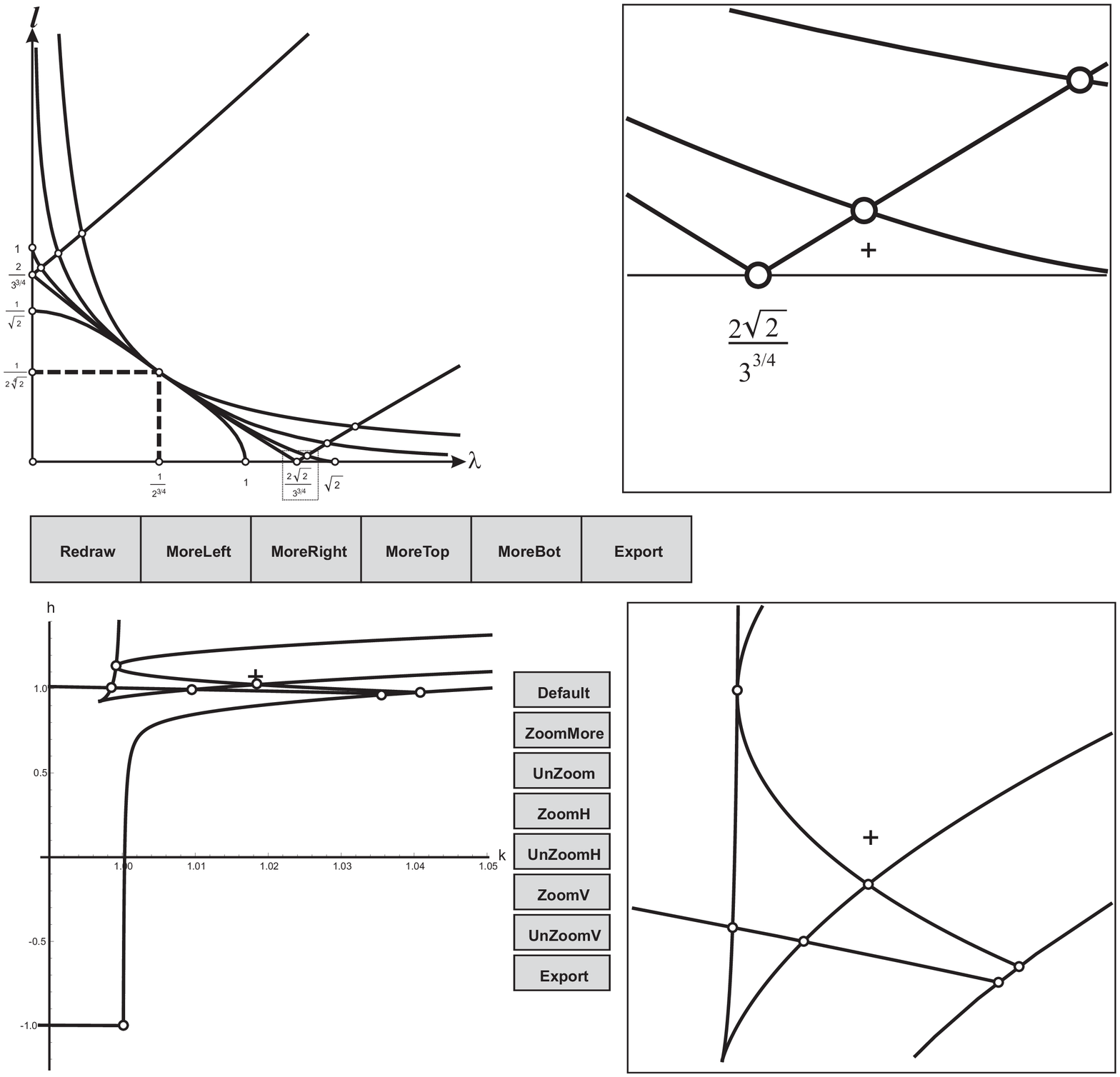}
\caption{Интерфейс электронного атласа.}\label{KhRy_fig5}
\end{figure}

%\sect{\label{sec6}Компьютерная реализация}
\begin{flushleft}
\noindent{\bf{\S\,5.\,Компьютерная реализация}}
\end{flushleft}

Бифуркационные кривые $(k(s),h(s))$ в сечении поверхностей $\bfa,\bfb$ в соответствии с допустимыми промежутками, установленными теоремой~\ref{teor-2}, строятся без каких-либо сложностей. В современных системах аналитических вычислений\footnote{При подготовке статьи авторы пользовались САВ <<Mathematica>>, лицензия ФГОУ ВПО ВАГС.} нахождение вещественных корней полиномиального уравнения в заданном промежутке выполняется встроенной функцией. Для поверхности $\bfc$ соответствующий алгоритм не должен, конечно, разбирать все варианты, указанные в теореме~\ref{teor-2} для полной строгости, а основываться на минимальном количестве логически возможных вариантов устройства промежутка $\mD_3(\la,\ell)$. Очевидно, $\mD_3(\la,\ell)$ есть полупрямая $s\in [s_{\rm min},+\infty)$ с возможным исключением двух <<лакун>> $ s \notin [\zeta^+,\eta_{23}]$ и $s \notin [\eta_{32}^-,\eta_{23}^+]$. Имеем
$$
s_{\rm min}=\left\{
\begin{array}{ll}
\mathop{\rm min}\nolimits \{ \zeta^-, \eta_{23} \}, & \ell \ls \ell_6 \\
\eta_{23}, & \ell > \ell_6
\end{array} \right.,
$$
а критерии существования лакун таковы:
\begin{equation}\label{eq6_1}
\begin{array}{l}
s \notin [\zeta^+,\eta_{23}] \quad \Longleftrightarrow \quad \la > \la_*, \quad \ell \in (\ell_4,\ell_6); \\
s \notin [\eta_{32}^-,\eta_{23}^+] \quad \Longleftrightarrow \quad \la > \la^*, \quad \ell \in [0, \ell_3).
\end{array}
\end{equation}

Нетрудно видеть, что условия \eqref{eq6_1} отражают все случаи, в которых допустимый промежуток для $s$ на сечении поверхности $\bfc$ несвязен.
Напомним, что через $\silaj$ обозначены бифуркационные кривые $\sila \cap \wsi_j$. На рис.~\ref{KhRy_fig4} приведены примеры диаграмм, в которых промежуток изменения параметра $s$ в $\ell$-сечении поверхности $\bfc$ состоит из одного, двух или трех отрезков. Звездочкой обозначены области значений интегралов, в которых решения системы \eqref{eq1_1} невозможны. Значение $\la$ достаточно велико, чтобы обеспечить неравенство $\ell_4<\ell_6<\ell_3$ (см. рис.~\ref{KhRy_fig3},$e$). При $\ell>\ell_3$ (рис.~\ref{KhRy_fig4},$a$) лакун нет, множество $\sthr$ состоит из одного отрезка (помечен цифрой 1 на рисунке). Когда $\ell$, уменьшаясь, пересекает $\ell_3$, точка возврата кривой $\sthr$ проходит через кривую $\sone$. Возникает вторая из лакун \eqref{eq6_1}, ограниченная точками из $\delta_2$. Две компоненты кривой  $\sthr$ отмечены цифрами 1, 2 на рис.~\ref{KhRy_fig4},$b$. При переходе $\ell$ вниз через $\ell_6$ возникают две точки касания поверхностей $\bfc$ и $\bfa$ на уровне $\{\ell=\cons\}$. Между ними образуется новый отрезок кривой $\sthr$ (отмечен цифрой 3 на рис.~\ref{KhRy_fig4},$c$). Здесь существует первая из лакун \eqref{eq6_1}, отвечающая недопустимому <<уголку>> справа от дуги $A_6-A_4$ на рис.~\ref{KhRy_fig3},$e$. Несмотря на тот факт, что в точке касания, имеющей большее значение $h$, значение $k$ быстро растет при дальнейшем уменьшении $\ell$, расстояние по $h$ между отрезками кривых $\sthr$ и $\sone$, ограниченных точками касаниями, остается малым.

На рис.~\ref{KhRy_fig5} показано окно программы, реализующей просмотр электронного атласа бифуркационных диаграмм $\sila$. В правой верхней четверти выводится разделяющее множество~$\mA$, определенное теоремой~\ref{teor-1}, с возможностью интерактивного выбора точки в плоскости $(\la,\ell)$ и последующим уточнением ее положения относительно разделяющих кривых в правой верхней четверти.
В нижней левой четверти строится точная бифуркационная диаграмма (подчеркнем, что возможность построения именно бифуркационной диаграммы, а не совокупности несущих ее кривых, реализована впервые и основана на аналитическом описании допустимых множеств).
В качестве начальной области построения берется область, содержащая образы всех критических точек ранга 0 и вырожденных критических точек ранга 1. Поскольку диаграмма неограничена, это область не всегда отображает все ее характерные черты (например, экстремальные значения интеграла $K$ на критических подсистемах). В связи с этим реализованы возможности <<подгонки>> отображаемой части диаграммы в целом. Справа от диаграммы показывается увеличенная область, центр которой указывает курсор, имеющий одно и то же положение как на самой диаграмме, так и в увеличенном фрагменте, что дает возможность его более тонкого позиционирования. Относительно выбранного положения курсора можно производить изменение масштаба изображения как по горизонтали, так и по вертикали. В частности, имея в виду дальнейшие задачи по визуализации бифуркаций торов Лиувилля в фазовом пространстве, основанные на алгоритмах, разработанных и примененных в работах \cite{RyabRCD, RyabDis}, можно при заданном $\ell$ обеспечить попадание точки $(k,h)$ в любую компоненту множества $\bbI\backslash \Sigma$, в том числе и в упомянутую выше "сверхузкую"\, область, порожденную кривой касания поверхностей $\bfa$ и $\bfc$. Также предоставлена возможность экспорта изображений фрагментов разделяющего множества и бифуркационных диаграмм.

Авторы выражают искреннюю признательность
профессору М.\,П.~Харламову за предложенный конструктивный подход, полезные советы
и постоянное внимание к работе.
%----------------------------------------------
\vspace{3ex}

%\clearpage

\small

\makeatletter \@addtoreset{equation}{section}
\@addtoreset{footnote}{section}
\renewcommand{\section}{\@startsection{section}{1}{0pt}{1.3ex
plus 1ex minus 1ex}{1.3ex plus .1ex}{}}

{ %\scriptsize

\renewcommand{\refname}{{\rm\centerline{СПИСОК ЛИТЕРАТУРЫ}}}

}

\titleeng


\begin{thebibliography}{99}

\re\bibitem{PVLect} Харламов~П.\,В. Лекции по
динамике твердого тела.~--- Новосибирск: Изд-во
НГУ, 1965.~--- 221~с.


\re\bibitem{PVMtt71} Харламов~П.\,В. Один
случай интегрируемости уравнений движения
твердого тела, имеющего неподвижную точку~//
Механика твердого тела.~---  1971.~--- №~3.~---
С.~57--64.



\re\bibitem{Yeh1} Yehia~H. New integrable cases
in the dynamics of rigid bodies~// Mech. Res.
Commun.~---  1986.~--- V.~13.~--- №~3.~---
P.~169--172.


\re\bibitem{Gash1} Гашененко~И.\,Н. Новый класс
движений тяжелого гиростата~// Докл. АН СССР.
~---  1991.~--- Т.~318.~--- №~1.~--- С.~66--68.

\re\bibitem{Ryab2} Рябов~П.\,Е. Некоторые
случаи вырождения переменных в одной задаче о
движении твердого тела вокруг неподвижной
точки~/ ВолГУ.~--- Волгоград, 1991.~---
9~с.~--- Деп. в ВИНИТИ. 20.06.1991,
\No~3660-В91.


\re\bibitem{Gash3} Гашененко~И.\,Н. Один случай
интегрируемости уравнений движения гиростата~//
Механика твердого тела.~---  1992.~---
№~24.~--- С.~1--4.

\re\bibitem{Gash4} Гашененко~И.\,Н.
Бифуркационное множество задачи о движении
гиростата, подчиненного условиям Ковалевской~//
Механика твердого тела.~---  1995.~---
№~27.~--- С.~31--35.

\re\bibitem{Ryab1} Рябов~П.\,Е. О вычислении
бифуркационного множества в случае
Кова\-лев\-ской-Яхьи~// Механика твердого
тела.~---  1995.~--- №~27.~--- С.~36--40.

\re\bibitem{Ryab5} Рябов~П.\,Е. Перестройки
бифуркационного множества в обобщенной задаче
С.В.\,Ко\-ва\-лев\-ской~/ ВолГУ.~--- Волгоград,
1996.~--- 7~с.~--- Деп. в ВИНИТИ. 19.02.1996,
\No~884-В96.


\re\bibitem{Ryab6} Рябов~П.\,Е. Об одном
свойстве бифуркационных кривых~/ ВолГУ.~---
Волгоград, 1996.~--- 16~с.~--- Деп. в ВИНИТИ.
15.01.1996, \No~1954-В96.

\re\bibitem{Ryab7} Рябов~П.\,Е. Бифуркационные
множества в задаче о движении твердого тела
вокруг неподвижной точки~// Вестник ВолГУ.~---
1996.~--- №~1.~--- С.~41--49.

\re\bibitem{Gash2} Гашененко~И.\,Н.
Бифуркационное множество в задаче о движении
тяжелого гиростата при условиях Ковалевской~//
Доповiдi НАН Украины.~---  1997.~--- №~2.~---
С.~60--62.

\re\bibitem{RyabRCD} Рябов~П.\,Е.,
Харламов~М.\,П. Бифуркации первых интегралов в
случае Ковалевской-Яхьи~// Regular and Chaotic
Dynamics.~---  1997.~--- Т.~2.~--- №~2.~---
С.~25--40.


\re\bibitem{FomGr} Фоменко~A.\,Т. Симплектическая топология вполне интегрируемых
гамильтоновых систем~// УМН.~--- 1989.~--- Т.~44.~--- №~1.~--- С.~145--173.


\re\bibitem{Gash5} Гашененко~И.\,Н.
Интегральные многообразия и топологические
инварианты одного случая движения гиростата~//
Механика твердого тела.~---  1997.~---
№~29.~--- С.~1--7.

\re\bibitem{GashDis} Гашененко~И.\,Н.
Инвариантные многообразия и множества
допустимых скоростей в динамике твердого тела:
дис. \ldots д-ра физ.-матем. наук.~/  ИПММ
НАНУ.~--- Донецк, 2008.~--- 300~с.


\re\bibitem{RyabDis} Рябов~П.\,Е.
Бифуркационное множество задачи о движении
твердого тела вокруг неподвижной точки в случае
Ковалевской-Яхьи: дис. \ldots канд. физ.-матем.
наук.~/ МГУ.~--- Москва,  1997.~--- 143~с.

\re\bibitem{ReySem} Рейман~А.\,Г.,
Семенов-Тян-Шанский~М.\,А. Лаксово
представление со спектральным параметром для
волчка Ковалевской и его обобщений~// Функц.
анализ и его приложения.~--- 1988.~---
Т.~22.~--- №~2.~--- С.~87--88.


\re\bibitem{Kh361} Харламов~М.\,П. Области
существования критических движений обобщенного
волчка Ковалевской и бифуркационные
диаграммы~// Механика твердого тела.~---
2006.~--- №~36.~--- С.~13--22.


\re\bibitem{KhKhSh40} Харламов~М.\,П.,
Харламова~И.\,И., Шведов~Е.\,Г. Бифуркационные
диаграммы на изоэнергетических уровнях
гиростата Ковалевской--Яхья~// Механика
твердого тела.~--- 2010.~--- №~40.~---
С.~77--90.


\re\bibitem{KhRCD05} Kharlamov~M.\,P.
Bifurcation diagrams of the Kowalevski top in two constant fields~// Regular and Chaotic Dynamics.~--- 2005.~--- V.~10, №~4.~--- P.~381--398.

\re\bibitem{KhND} Харламов~М.\,П. Критические
подсистемы гиростата Ковалевской в двух
постоянных полях~// Нелинейная динамика.~---
2007.~--- Т.~3.~--- №~3.~--- С.~331--348.

\re\bibitem{RyabUdgu} Рябов~П.\,Е.
Аналитическая классификация особенностей
интегрируемого случая Ковалевской-Яхья~//
Вестник УдГУ.~--- 2010.~--- №~4.~--- С.~25--30.

\re\bibitem{Ryab37} Рябов~П.\,Е.
Алгебраические кривые и бифуркационные диаграммы двух интегрируемых задач~//
Механика твердого тела.~--- 2007.~--- №~37.~--- С.~97--111.

\end{thebibliography}
\end{document}